\documentclass[english,aps,prl,superscriptaddress,floatfix,notitlepage,reprint,show pacs, iopart]{revtex4-2}
\usepackage[T1]{fontenc}
\usepackage[utf8]{inputenc}
\usepackage{physics}
\usepackage{natbib}
\usepackage{amsthm}
\usepackage{amssymb}
\usepackage{dsfont}
\usepackage{amsmath}
\usepackage{bm}
\makeatletter

\usepackage{braket}
\usepackage{txfonts} 
\usepackage{graphicx}

\usepackage[usenames,dvipsnames]{xcolor}
\usepackage[colorlinks=true,citecolor=Blue,linkcolor=RubineRed,urlcolor=Blue]{hyperref}
\usepackage{url}
\usepackage{tikz}
\usepackage{rotating} 
\usepackage{orcidlink}
\date{\today}

\makeatother

\begin{document}
\title{ Dynamical Crossover in Landau–Zener Tunneling in Dissipative Rydberg Lattices}
\author{Suvechha Indu \orcidlink{0009-0004-6334-0691}}
\thanks{Corresponding author: \href{suvechhaindu@gmail.com}{suvechhaindu@gmail.com}}
\address{Department of Physics, University of Calcutta, $92$ A. P. C. Road, Kolkata $700009$, India}
\author{Raka Dasgupta  \orcidlink{0000-0003-2148-4641}}
\email{rdphy@caluniv.ac.in}
\address{Department of Physics, University of Calcutta, $92$ A. P. C. Road, Kolkata $700009$, India}

\begin{abstract}
In this work, we investigate the excitation dynamics of a Rabi-coupled dissipative Rydberg
lattice with a time-dependent detuning. The system is analyzed using (i) a Lindblad master
equation within a mean-field approximation and (ii) an effective non-Hermitian Hamiltonian
framework. While the mean-field approach captures the emergence of an antiferromagnetic
order in the Rydberg excitation profile, the non-Hermitian description provides direct insight
into the complex energy spectrum and its avoided crossings, which govern the Landau–Zener dynamics. We identify a regime in which the sublattice population imbalance vanishes near the avoided crossing, resulting in identical Landau–Zener probabilities on the two sublattices. Beyond a critical effective blockade strength there is a dynamical crossover to another regime in which the sublattice population imbalance persists through the avoided crossing, giving rise to sublattice-dependent Landau–Zener probabilities. Furthermore, Rydberg interactions prolong the lifetime of Landau–Zener-induced excitations in the presence  of weak dissipation and strong Rabi coupling.  In contrast, for weak Rabi coupling, the Rydberg blockade inhibits excitation and suppresses the Landau–Zener transition probability.

\end{abstract}

\pacs{67.85.-d, 42.50.Gy, 3.67.-a, 03.67.Bg, 05.70.Fh, 05.90.+m, 37.10.Jk, 03.65.Yz, 03.65.Vf  }

\maketitle
\section{Introduction}
\label{intro}
The physics of open quantum systems is an immensely active area of research at present. While the physics of closed quantum systems follow relatively more established frameworks, open systems still involve ``open" questions. Also, there exists no single or universally accepted formalism to study the dynamics of these class of systems.  Solution of Lindblad \cite{carmichael1993open, breuer2002theory, manzano2020short} Master's equation \cite{carmichael2013statistical} is probably the most widely used template for this. Another option that has recently been popular is the use of non-Hermitian Hamiltonians \cite{roccati2022non, ashida2020non}. 

Traditional quantum mechanics involves Hermitian operators with real eigenvalues that automatically ensure the conservation of probability amplitude, applicable in closed systems. In reality, the systems are often connected to environment, and hence become open quantum systems. These open systems can be modeled with non-Hermitian Hamiltonians, and their complex eigenvalues encode both energies and decay rates. For example, radioactive nuclear decay can be mapped onto a non-Hermitian framework \cite{siegert1939derivation, majorana2006scattering, feshbach1958unified}, while a three-level atomic system with spontaneous decay can be represented by a non-Hermitian XY model \cite{lee2014heralded}. Non-Hermitian systems exhibit several intriguing properties like the existence of exceptional points \cite{longstaff2019nonadiabatic, melanathuru2022landau}, and non-Hermitian skin effects \cite{bergholtz2021exceptional, kawabata2021topological}. 

Non-Hermitian Hamiltonians arise in a variety of contexts. One important class consists of Pseudo-Hermiticity \cite{mostafazadeh2002pseudo} (for example, the PT-symmetric Hamiltonians \cite{bender2005introduction,bender2007making,tong2023adiabatic, ashida2020non, roccati2022non}), which possesses real spectra in the symmetry-unbroken phase. Another widely studied class comprises effective non-Hermitian Hamiltonians having complex eigenvalues \cite{ zhu2014pt, modak2021eigenstate, zhang2024pt, ohlsson2021transition,bender2017behavior,li2021spectral} describing dissipative open quantum systems \cite{halder2023properties}.

A standard Lindblad evolution can be decomposed into a non-Hermitian evolution \cite{roccati2022non} under an effective Hamiltonian and a quantum-jump term \cite{carmichael1993open}. Neglecting the latter results in  the effective non-Hermitian description.  One added advantage of modeling dissipative systems using non-Hermitian Hamiltonians is that the energy spectrum can be obtained directly from the Hamiltonian itself. In the master-equation framework, the system Hamiltonian is Hermitian, while dissipation is incorporated through additional Lindblad terms.  While this approach successfully captures the system dynamics, extracting effective energy spectra is not always straightforward. In contrast, under the non-Hermitian mapping, the Hamiltonian matrix can be directly diagonalized to obtain complex eigen-energies. This makes it possible to identify avoided level crossings and connect them to phenomena such as Landau–Zener (LZ) transitions \cite{melanathuru2022landau, longstaff2019nonadiabatic}.

In recent years, the LZ transition has been studied for dissipative two-level \cite{javanbakht2015dissipative, saito2007dissipative, wubs2006gauging, dai2025dissipative} and three-level \cite{militello2019three} systems, connected to external bath. The formalisms employed include Lindblad Master equation \cite{arceci2017dissipative}, numerical path-integral approach \cite{javanbakht2015dissipative, nalbach2009landau, chen2020landau}, discrete-time stochastic shr\"odinger equation \cite{barra2016dissipation}, Floquet-based approach \cite{bonifacio2020landau, ferron2016dynamic}, etc. However, in most of these works, the focus is  on the population dynamics only, and the corresponding energy spectra is not readily accessible. As a result, although the LZ transition has been reported to exhibit intriguing features, their connection to the underlying energy landscape has often remained unexplored.  This is precisely where the non-Hermitian approach has a clear added advantage \cite{militello2019three, torosov2017pseudo} as it provides direct access to the energy spectra and can serve as a bridge between LZ dynamics and the underlying energy landscape.

The non-Hermitian framework, however, has certain limitations. For example, the equivalence of non-Hermitian and Liouvillian dynamics can be perfectly established for semiclassical limits only \cite{minganti2019quantum}. Similarly, the exact solution of a two site decay system is well-represented by non-Hermitian dynamics only in the weak-coupling and the singular coupling limits, while the Lindblad dynamics is accurate \cite{monkman2026limits} for arbitrary parameter strengths. It has also been shown that the mixed-state open quantum dynamics with a Lindbladian can be mapped to a pure-state
non-Hermitian dynamics upon a postselection of measurement outcomes \cite{chaduteau2026lindbladian}. In some cases, a combination of the non-Hermitian dynamics and the Lindblad master equation is useful \cite{zloshchastiev2014comparison}, too.

Thus, the Lindblad and the non-Hermitian frameworks are sort of complementary to each another. There is a clear trade-off: The Lindblad master equation provides a more accurate description of the population dynamics (as it also accounts for the quantum-jump processes), while the non-Hermitian formulation offers direct access to the complex eigen-energy spectra. 

Rydberg lattices have drawn much attention in recent times. Their strong dipole–dipole interactions lead to the Rydberg blockade, which prevents simultaneous excitation of nearby atoms. This leads to the emergence of an antiferromagnetic order \cite{ebadi2021quantum, scholl2021quantum} across the lattice, and enables the generation of entangled states \cite{levine2018high, evered2023high}. Moreover, this antiferromagnetic order has very interesting dynamical properties: they do not thermalize \cite{bernien2017probing} in smaller timescales, and shows scaling behaviors \cite{keesling2019quantum} near quantum critical points. Given that Rydberg lattices host a wide range of fascinating quantum phenomena, it is essential to investigate their spectral properties and population dynamics, with a focus on the role of the Rydberg blockade mechanism.

In a recent work, the entanglement dynamics of ultracold Rydberg atoms under Landau–Zener (LZ) sweeps was investigated \cite{varghese2023maximally}. In absence of dissipation, the LZ transition probability was directly obtained here from the coherent Hamiltonian dynamics, and dissipation was incorporated later through the Lindblad form. They observed that a Rydberg atom pair evolves periodically through various maximally entangled states. The thrust here was on population transfer and entanglement measures, and not on the spectral properties.

In the present work, we formulate a non-Hermitian description of a dissipative Rydberg system, which allows us to directly investigate the roles of complex eigenenergies, avoided crossings, and exceptional points, and to reveal new aspects of the LZ dynamics. Considering a time-dependent detuning parameter, we calculate the energy spectra in a regime where the quantum-jump term is negligible, to ensure that the non-Hermitian evolution faithfully captures the dynamical behavior. We establish a direct connection between characteristic features of the system dynamics and the underlying energy spectra. We find that the emergence of antiferromagnetic order corresponds to the appearance of  avoided crossings in the energy-level structure. We further show that the Rydberg interaction strength controls the lifetime of LZ-induced transitions, which can be inferred from the number of avoided crossings. Finally, we identify a dynamical crossover in the Landau–Zener transition probabilities and show that it can be understood in terms of the underlying energy landscape.

The paper is organized as follows. In Sec.~\ref{Heff} we present the model Hamiltonian and an outline for solving it via (i) Linblad master equation, and (ii) non-Hermitian Hamiltonian.  Sec.~\ref{dynamics} contains the phase plots of the system in terms of the population distribution, and enables us to identify the parameter regime for which both the approaches yield similar results. In Sec.~\ref{spectra}, we show emergence of loops and branches in the eigen-energy spectra upon variation of the Rydberg strength and the avoided crossing shifts from the imaginary spectra to the real spectra upon parameter variation. The dependency of Landau-Zener transition probability on dissipation and Rydberg strength is accommodated in Sec.~\ref{LZ}. Sec.~\ref{crossover} contains, crossover from a region, where, LZ transition probabilities for the two sublattices are overlapped to a region, where they are distinct. Finally in Sec.~ \ref{conclusion}, we conclude our work on the non-Hermitian Rydberg system.

\section{Construction of an Effective Hamiltonian }
\label{Heff}

Here we study a two-level, non-equilibrium system with ultracold Rydberg atoms in an optical lattice. The ground state is denoted by $|G\rangle$, and the excited state is denoted by $|E\rangle$. The atoms can decay from the higher energy state to the ground state, and that is the dissipation rate $\gamma$ here. The detuning between the two states can be made time-varying using an external time-dependent laser field or magnetic field.

The Hamiltonian of the system is 
\begin{equation}
    \begin{split}
        H  = & \sum_{j} \Big(-\Delta (t)|E\rangle \langle E|_{j} + {{\Omega}\over{2}}(|E\rangle \langle G|_{j} + |G\rangle \langle E|_{j})\Big)\\ &  + V\sum_{<j k>} |E\rangle \langle E|_{j} \otimes |E\rangle \langle E|_{k}
    \end{split}
    \label{HHamiltonian}
\end{equation}
Here, $\Delta (t)$ is time dependent detunning parameter, $\Omega$ is Rabi coupling, $V$ is nearest neighbor Rydberg interaction. Also, $j, k$ are the site indices.

The dynamics is obtained by using this Hamiltonian (which is Hermitian by construction) in the Lindblad master equation.

\begin{equation}
\dot{\rho} = -i[H, \rho] + \gamma\sum_i \Big(\hat{L_i} \rho \hat{L_i^\dagger} - {{1}\over{2}}\{ \hat{L_i^\dagger} \hat{L_i},\rho \}\Big)\\
\label{lindblad}
 \end{equation}
Here $\rho$ is density matrix of the system with elements $\rho_{GG}$, $\rho_{GE}$, $\rho_{EG}$, and $\rho_{EE}$, while $\hat{L} = |G\rangle \langle E|$. In Eq. \ref{lindblad}, the Hamiltonian yields the coherent dynamics, while the Lindblad  terms generate the dissipative evolution. This can be recast as  \cite{roccati2022non}:
 \begin{equation}
     \dot{\rho}=-i(H_{\mathrm{eff}} \rho - \rho H_{\mathrm{eff}}^{\dagger})+\gamma\sum_i \hat{L_i} \rho \hat{L_i^\dagger}
     \label{nh1}
 \end{equation}
 $H_{\mathrm{eff}}$ has the general form 
 \begin{equation} 
 H_{\mathrm{eff}}=H-{{i}\over{2}}\gamma\sum\limits_i\hat{L_i^\dagger}  \hat{L_i}
 \end{equation}
For our system, this translates to 
\begin{equation}
    \begin{split}
        H_{\mathrm{eff}}  = & \sum_{j} \Big(-({{i \gamma}\over{2}}+\Delta(t) )|E\rangle \langle E|_{j} + {{\Omega}\over{2}}(|E\rangle \langle G|_{j} + |G\rangle \langle E|_{j})\Big)\\ &  + V\sum_{<j k>} |E\rangle \langle E|_{j} \otimes |E\rangle \langle E|_{k}
    \end{split}
    \label{NHHamiltonian}
\end{equation}

The second term in Eq. \ref{nh1} is the quantum jump term. If this term is dropped, the system evolves under the effective non-Hermitian Hamiltonian and the energy eigen-spectra can be extracted as well. However, neglecting this term (In our model, this is equal to $\sum_j\gamma\rho_{EE,j}|G\rangle\langle G|$ ) would mean, one is evolving a non-unitary conditional state under the assumption that  no explicit spontaneous decay has occurred within that time-step of the time-discretized evolution process. Thus, all quantum trajectories are not sampled in defining the dynamics \cite{le2024entanglement}. Moreover, in our treatment, the  density matrix is assumed to remain normalized, i.e. $\rho_{EE,j}+\rho_{GG,j}$ at all instant even though the no-jump evolution is not trace preserving. Thus the non-Hermitian description only generates an approximate dynamics. Nevertheless, this is an useful approximation, valid throughout a wide range in the parameter space, as we shall see in Sec. \ref{compare_theta}.

We consider a mean-field approximation to study the system \cite{lee2011antiferromagnetic, indu2026different}. With this, the $j^{th}$ Rydberg interaction term in Eq. \ref{NHHamiltonian}, $V |E\rangle\langle E|_j \otimes\sum_k |E\rangle\langle E|_k$ is approximated as, $V|E\rangle\langle E|_j \otimes\sum_k \rho_{EE,k}$. Hence the non-Hermitian Hamiltonian can be written in matrix form as:
\begin{equation}
H_{\mathrm{eff}}=
\begin{pmatrix} 
0 & \Omega\over2 \\ \Omega\over2 &
-({{i \gamma}\over{2}}+ \Delta(t) )+V \sum_k \rho_{EE,k} 
\end{pmatrix}
\label{matrix}
\end{equation}

Also, the population imbalance between the ground state $|G\rangle$ and the excited state $|E\rangle$ is defined as $\omega$ \cite{indu2026different, lee2011antiferromagnetic}. i.e. $\omega_k=\rho_{EE,k} - \rho_{GG,k} $. Combining this with the normalization of the density matrix, gives the relations:

\begin{equation}
    \rho_{EE,k} = {{1+\omega_k
    }\over{2}}
    \label{rhoe}
\end{equation}
\begin{equation}
    \rho_{GG,k} = {{1-\omega_k
    }\over{2}}
    \label{rhog}
\end{equation}
So $\omega_k$ essentially is a measure that reflects the number of Ryderg excitations in $k^{th}$ site. 

In Sec. \ref{dynamics}, we compare between phase plots (in terms of the Rydberd excitations) obtained using Lindblad and non-Hermitian frameworks, and identify the parameter regime where the later can be considered as a valid approximation.

\section{Dynamics}
\label{dynamics}
As discussed before, constructing the non-Hermitian effective Hamiltonian enables one to have a direct insight into the complex eigen-energies. However, the representation is valid only when the quantum jump term is indeed negligible. So we first solve Eq. \ref{nh1} both with and without the jump terms, and from each approach, construct  order parameter plots. A comparison between the two helps us to identify the parameter regime within which the no-jump description correctly reproduces the qualitative behavior as obtained in the full solution. Then, in Sec. \ref{spectra} and \ref{LZ}, we study the spectral properties of the system in that regime.  
\subsection{System Parameters} 
We consider a detuning that is linear in time, i.e., $\Delta=v t$. We express all energies in units of a reference interaction strength
$V_0=C_6(n_0)/R_0^6$ corresponding to the Rydberg system (and set $\hbar =1$): where $C_6$ is the van der Waals interaction coefficient, $n_0$ is a reference principal quantum number, and $R_0$ is the lattice spacing. For example, with $^{87}\mbox{Rb}$ atoms and $S_{1/2}$ electrons, one can choose $n_0=50$
as the reference principal quantum number. Then, if the lattice spacing is roughly $10 \mu m$ in frequency units, $V_0=15 $ MHz. Just by changing the principal quantum number $n$, one can vary $V$:

\begin{table}[h]
\centering
\begin{tabular}{|c | c | }
\hline
$V/V_0$ & Corresponding $n$ \\
\hline
3  & 55 \\
\hline
5  & 58 \\
\hline
10 & 62 \\
\hline 
21 & 66 \\
\hline
\end{tabular}
\caption{Approximate mapping between interaction strength and principal quantum number for Rb Rydberg states, using $n_0=50$ as reference.}
\end{table}

This range of $n$ values as well as the lattice separation fall well within possible experimental ranges \cite{viteau2013rydberg, beguin2013direct, leung2014magnetic, singer2004suppression}. All other parameters ($\Omega$, $\gamma$ and $v$) are expressed in units of this reference $V_0$. In our calculations, we mostly use $\Omega \approx 0.3$ or 0.5 (corresponding to 4.5 MHz, and 7.5 MHz respectively in frequency units), well within the experimentally accessible range. 

The parameter $\gamma$ is chosen such that it can  capture both  spontaneous emission \cite{lee2019coherent} and additional experimental decoherence channels \cite{day2008dynamics}. Small values of $\gamma$ correspond to radiative decay, while larger values represent engineered  dissipation. For most of our calculations, we take $\gamma=0.001$ to describe weak dissipation (spontaneous decay) and $\gamma=0.4$ to describe strong dissipation, achieved through artificial coupling to rapidly decaying states, or via optical pumping. 

\subsection{Phase Plot}
The Hamiltonian (Eqs. \ref{HHamiltonian}, \ref{NHHamiltonian}) produces a bipartite lattice system in terms of the Rydberg population, i.e., one sublattice contains high Rydberg-excited population and another contains a lower one due to the presence of Rydberg blockade. In certain regions in the  parameter space, the Rydberg excitations for the two sublattices ( reflected in $\omega_1, \omega_2$) are equal which is called the uniform phase \cite{indu2026different, lee2011antiferromagnetic}. In some other regions, $\omega_1$ and $\omega_2$ are different, and that is the non-uniform phase \cite{indu2026different, lee2011antiferromagnetic}.

Depending upon the stability of the $\omega_1$ and $\omega_2$ solutions, this non-uniform phase has two subcategories. If   $\omega_1,\omega_2$ represent unequal but stable populations, that is termed as the antiferromagnetic phase \cite{indu2026different, lee2011antiferromagnetic}. On the other hand, if $\omega_1$ and $\omega_2$ are unequal but unstable (oscillating in time), it is called the oscillatory phase \cite{indu2026different, lee2011antiferromagnetic}. 

To characterize different phases of the system, an order parameter $\theta$ can be constructed as $\theta=|\omega_1 - \omega_2|/2$ \cite{indu2026different}. A zero $\theta$ indicates an uniform phase, a non-zero stable $\theta$ corresponds to the antiferromagnetic phase, while a fluctuating and non-zero $\theta$ marks the oscillatory phase \cite{indu2026different, lee2011antiferromagnetic}.

\subsection{Comparison of the phase plots: Hermitian vs non-Hermitian}
\label{compare_theta}

\begin{figure}
\begin{center}
\includegraphics[width=0.49\linewidth]{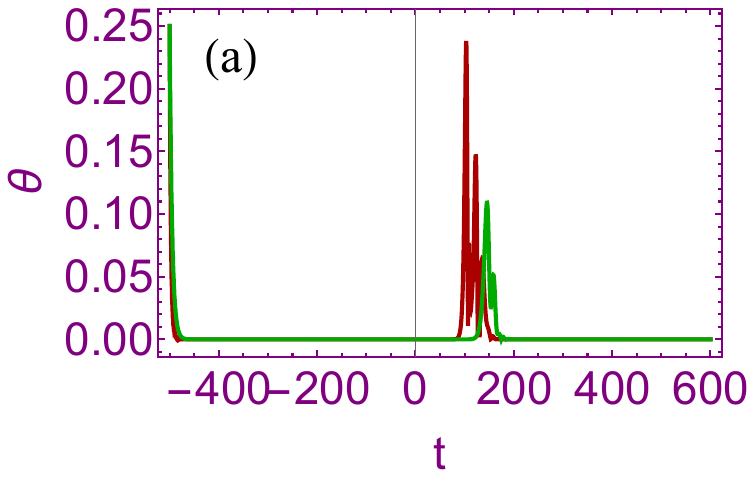}
\includegraphics[width=0.49\linewidth]{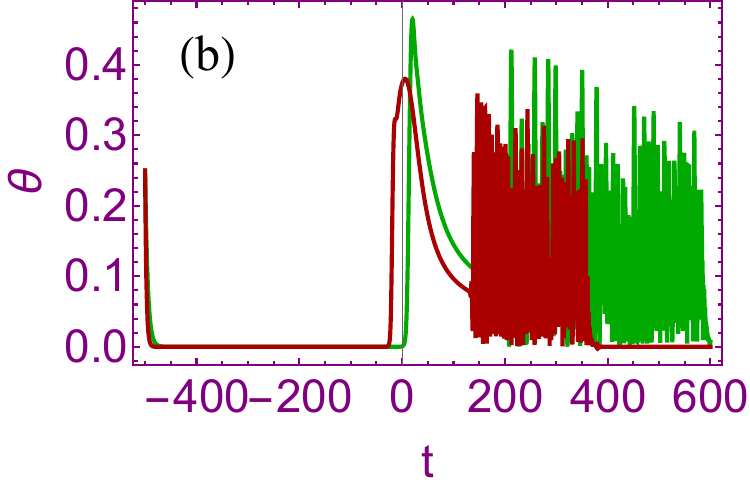}
\includegraphics[width=0.49\linewidth]{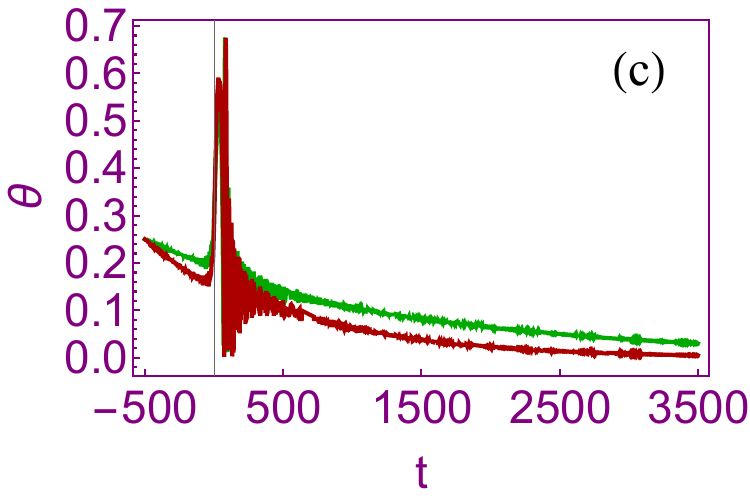}
\includegraphics[width=0.49\linewidth]{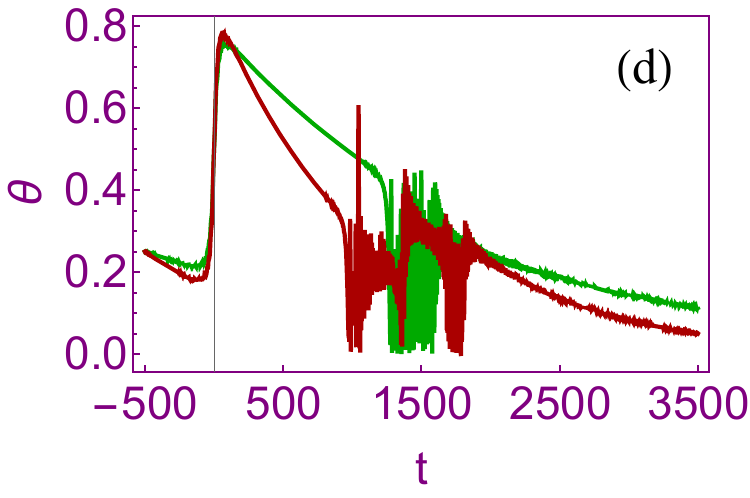}
\caption{Order parameter ($\theta$) vs time ($t$) plot (phase plot) extracted from Linblad system (maroon) and non-Hermitian system (green). For moderate dissipation, the parameters are $\Omega=0.5, \gamma =0.4, v = 0.01$ with (a) $V =4$, (b) $V =30$. For a higher V, the peaks show a better overlap. For low dissipation, the parameters are $\Omega=0.3, \gamma =0.001, v = 0.01$, with (c) $V=1$ (d) $V=30$. Here, the phase plots from the two approaches match reasonable well.}
\label{peak}
\end{center}
\end{figure}

 To construct the phase diagram, we plot the order parameter $\theta$ as a function of time $t$. In Fig. \ref{peak}, we present the phase diagrams corresponding to (i) the original Hermitian Hamiltonian clubbed with the Lindbladian (Maroon), (ii) the equivalent non-Hermitian Hamiltonian (Green). While the former treatment accommodates the quantum jump, the later approach neglects it. 

 We consider two distinct cases: (I) moderate dissipation ($\gamma = 0.4$) and (II) low dissipation ($\gamma = 0.001$). For (I), we choose the parameters $\Omega = 0.5, v = 0.01, \gamma = 0.4$ and find uniform phase from $V = 0$ to $V=3$ from both approaches. From $V=4$, both phase plots shows non-uniform phase (Fig. \ref{peak}(a)). As $V$ increases, the area of the oscillatory region corresponding to the non-Hermitian Hamiltonian increases faster than the Linblad system (Fig. \ref{peak}(b)). However, the gap between the peaks of the non-uniform phases decreases (Fig. \ref{peak}(b)) with an increasing $V$, making the matching more robust.

For (II), we take $\Omega = 0.3, v = 0.01, \gamma = 0.001$ and find that the entire non-uniform regions as obtained from non-Hermitian and Lindblad approaches match  considerably, from $V=1$ (Fig. \ref{peak}(c)) to $V=30$ (Fig. \ref{peak}(d)). 

\section{Excitation spectra of non-Hermitian Hamiltonian}
\label{spectra}

\subsection{Loops and branches}

\begin{figure}
\begin{center}
\includegraphics[width=0.49\linewidth]{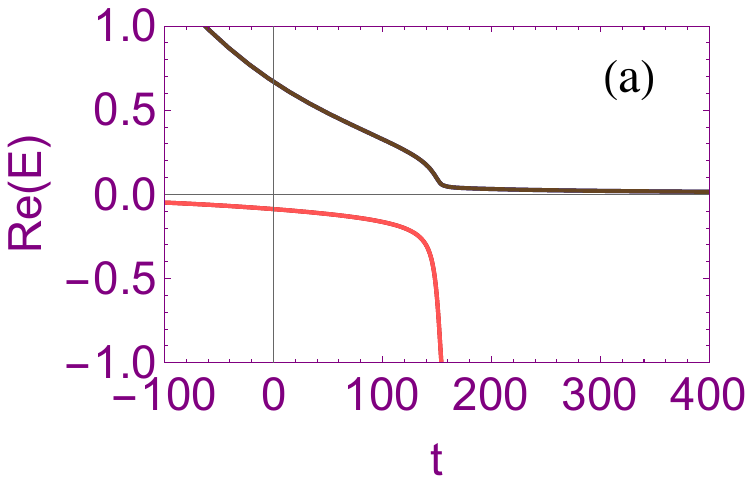}
\includegraphics[width=0.49\linewidth]{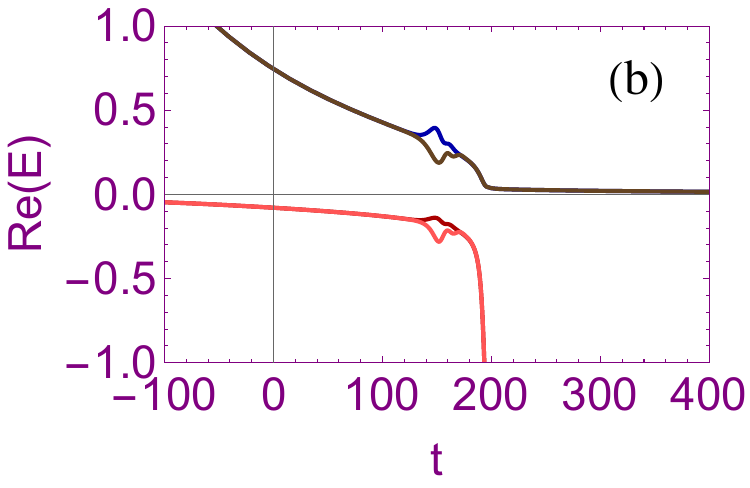}
\includegraphics[width=0.49\linewidth]{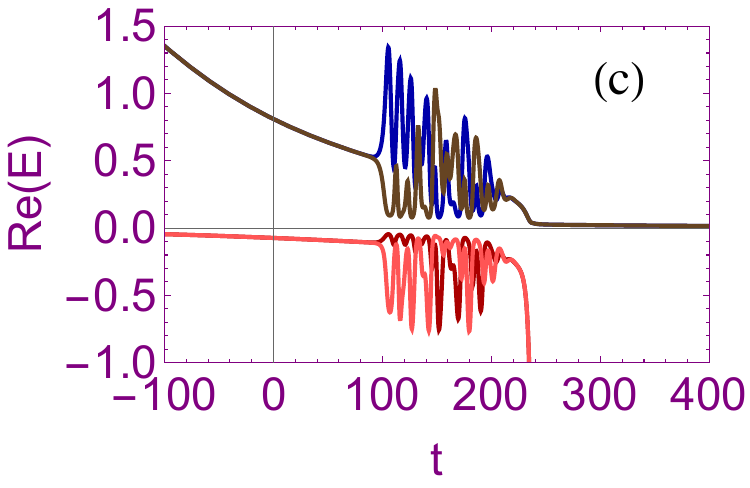}
\includegraphics[width=0.49\linewidth]{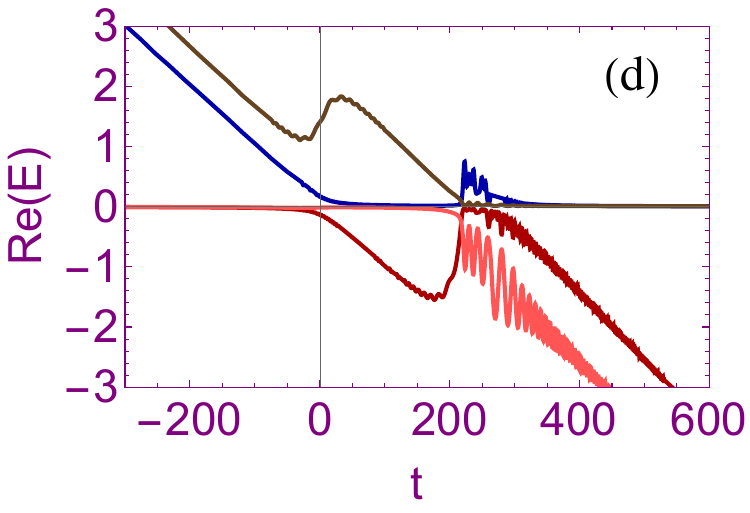}

\caption{ Real part of the energy eigen-spectra of the system showing avoided crossings. For $\gamma = 0.4,\Omega =0.5,  v = 0.01$ and (a) $V=3$: there is no loop in the spectra, (b) $V=4$: a loop appears near the avoided crossing, (c) $V=5$: oscillations appear in the loop. For $\gamma=0.001, \Omega=0.3,  v=0.01$, (d) $V=4$: there are sublattice-dependent branches in the spectra. Here, the colors, blue, brown, pink and maroon represent four branches of the eigen-energy spectra of the two state-two sublattice system. }

\label{spectrareal}
\end{center}
\end{figure}

We first study the real part of the energy eigen-spectra of the non-Hermitian Hamiltonian. Since there are two energy levels and two sublattices involved, there should, in principle be four energy branches in the real part of the eigen-energy. We find that for moderate  dissipation ($\gamma=0.4$) and small Rydberg interaction strength ($V=3$), the energies corresponding to the two-sublattices match exactly (Fig. \ref{spectrareal}(a)), and only two distinct branches are visible. As $V$ increases to $4$, a small loop is created near the avoided crossing, indicating an antiferromagnetic order (Fig. \ref{spectrareal}(b)). Further increment of $V$ to $V=5$ introduces an oscillation in the loop (Fig. \ref{spectrareal}(c)). 
At low dissipation limit $(\gamma=0.001)$, there is no such loop but four discrete branches are visible (Fig. \ref{spectrareal}(d)) even for $V=3$.

\subsection{Avoided Crossings in Real and Imaginary Spectra }

We observe that for a certain  parameter range, an avoided crossing appears in the real part of the energy eigen-spectra and for another range, the avoided crossing is present in the imaginary part. At a critical $\Omega/\gamma$ ratio, there are true crossings in both real and imaginary spectra. below the critical ratio, avoided crossing is in the imaginary spectrum, and above the ratio, the avoided crossing is in the real spectrum. For our model, this critical value is $\Omega/\gamma =1/2$. 

\begin{figure}
\begin{center}
\includegraphics[width=0.49\linewidth]{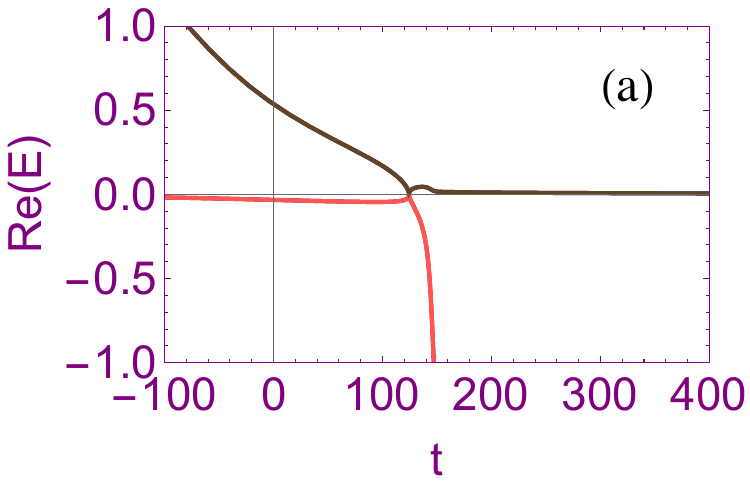}
\includegraphics[width=0.49\linewidth]{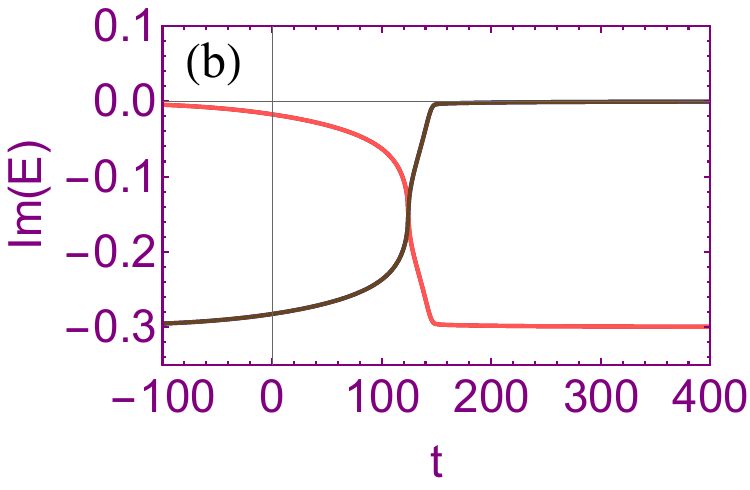}
\caption{Real (Re(E)) and imaginary (Im(E)) parts of the eigen-energy spectra at $\gamma =0.6, V =5, v = 0.01, \Omega=0.3$. (a) Real spectrum, (b) imaginary spectrum. Here, $\Omega/ \gamma = 1/2$, which is the critical $\Omega/ \gamma $ ratio, where, both the spectra undergo real crossings. }
\label{criticalomega}
\end{center}
\end{figure}

\begin{figure}
\begin{center}
\includegraphics[width=0.49\linewidth]{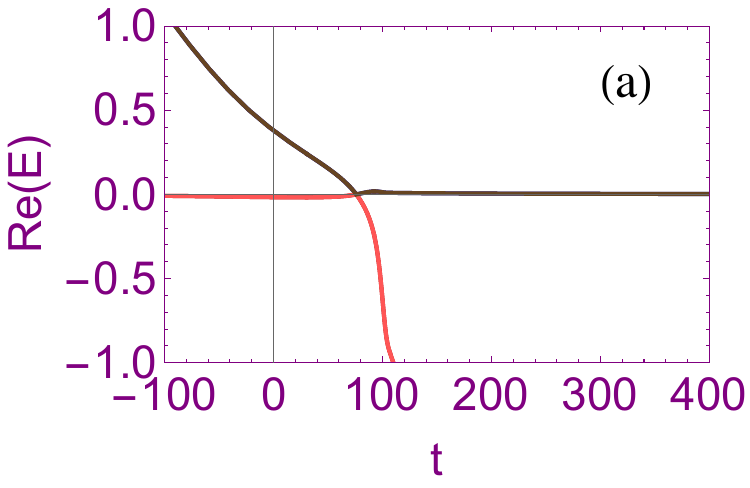}
\includegraphics[width=0.49\linewidth]{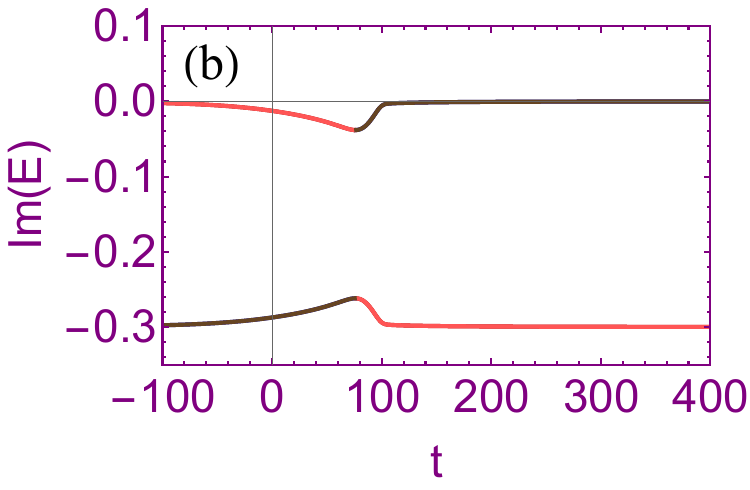}
\caption{Real (Re(E)) and imaginary (Im(E)) part of the eigen-energy spectra at $\gamma =0.6, V =5, v = 0.01, \Omega=0.2$. (a) Real spectrum, (b) imaginary spectrum. Here, $\Omega/\gamma$ ratio is below the critical value, where, avoided crossing is in the imaginary spectrum and the real spectrum contains true crossing. }
\label{belowcriticalomega}
\end{center}
\end{figure}

\begin{figure}
\begin{center}
\includegraphics[width=0.49\linewidth]{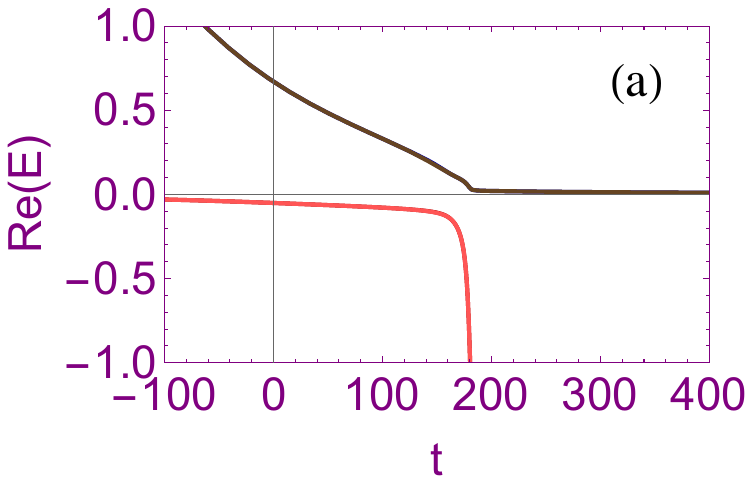}
\includegraphics[width=0.49\linewidth]{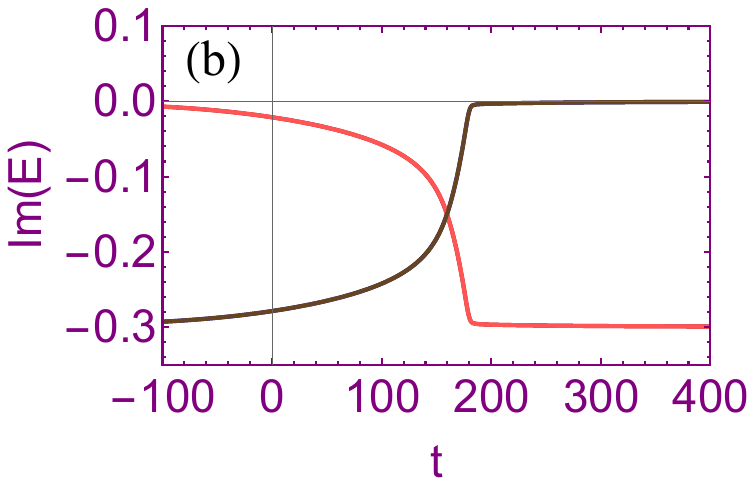}
\caption{Real (Re(E)) and imaginary (Im(E)) part of the eigen-energy spectra at $\gamma =0.6, V =5, v = 0.01, \Omega=0.4$. (a) Real spectrum, (b) imaginary spectrum. Here, $\Omega/\gamma$ ratio is above the critical value, where, avoided crossing is in the real spectrum and, the imaginary spectrum contains real crossing.}
\label{abovecriticalomega}
\end{center}
\end{figure}

From the effective Hamiltonian matrix in Eq. \ref{matrix}, the exceptional point of the system corresponds to:
\begin{equation}
\Big(\frac{i \gamma}{2}+vt - \frac{V(1+\omega(t))}{2}\Big)^2+\Omega^2=0
\end{equation}
where $\omega(t)=\omega_1(t)=\omega_2(t)$ in the uniform region. This means, the EP appears only for a specific time that is solution of the self-consistent equation:
\begin{equation}
t=\frac{V(1+\omega(t))}{2v}
\end{equation} 
with the condition $\Omega/\gamma=1/2$. At this point, there is real crossing in both real and imaginary spectra (Figs. \ref{criticalomega}(a), (b)). When the ratio is below this critical ratio ($\Omega/\gamma=1/4$ in Fig. \ref{belowcriticalomega}) the avoided crossing appears in the imaginary part of eigen spectra (Fig. \ref{belowcriticalomega}(b)), while the real part undergoes a real crossing (Fig. \ref{belowcriticalomega}(a)). In contrast, when the ratio is above this critical ratio ($\Omega/\gamma=3/4$ in Fig.   \ref{abovecriticalomega}), the avoided crossing appears in the real spectrum (Fig. \ref{abovecriticalomega}(a)) and the imaginary spectrum hosts a real crossing (Fig. \ref{abovecriticalomega}(b)). 

\section{Landau-Zener probability}
\label{LZ}
\subsection{Role of Dissipation }

\begin{figure}
\begin{center}
\includegraphics[width=0.49\linewidth]{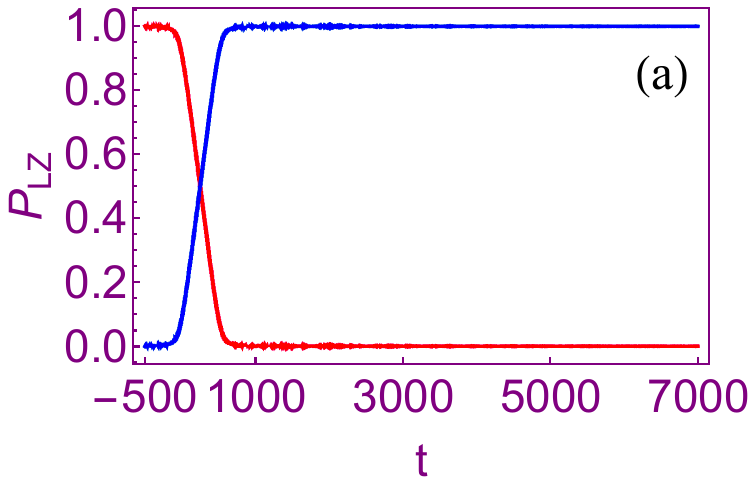}
\includegraphics[width=0.49\linewidth]{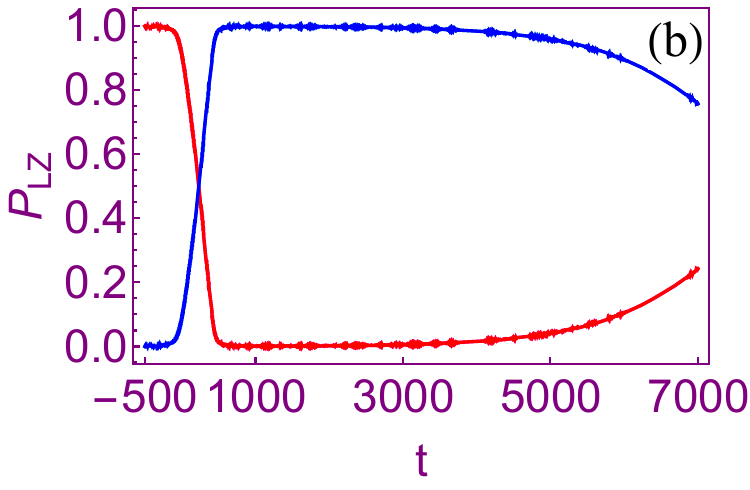}
\includegraphics[width=0.49\linewidth]{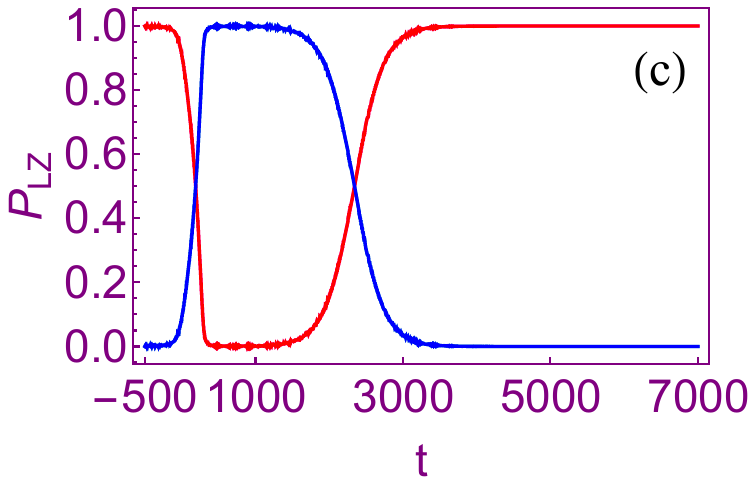}
\includegraphics[width=0.49\linewidth]{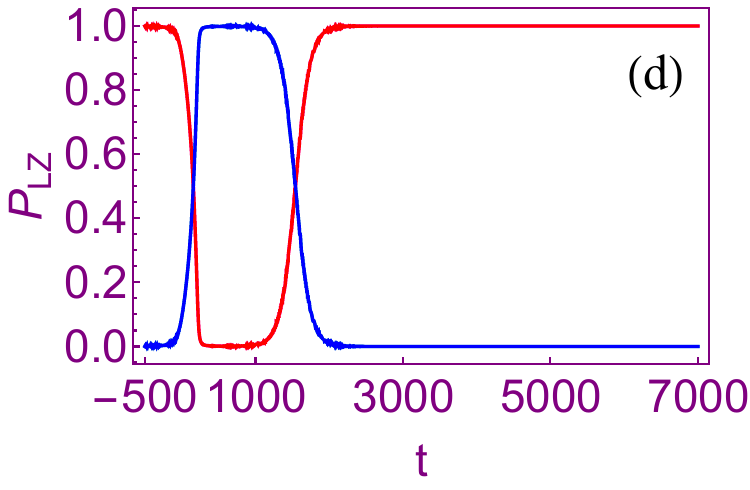}
\caption{Landau-Zener transition probability $P_{\mathrm{LZ}}$ at $\Omega =0.3, V =5, v = 0.01$ (a) $\gamma = 0$, (b) $\gamma = 0.001$. The excited state population survives from $t =248 $ to $t > 7000 $, (c) $\gamma = 0.005$: excited state population survives from $t =248 $ to $t = 2367 $, (d) $\gamma = 0.01$: the excited state population survives from $t =163 $ to $t = 1604 $. (We consider the window where the excited state occupation probability exceeds the ground state occupation probability ) Here, the red curve denotes occupation probability of the ground state population and blue color indicates the same for the excited state population.}
\label{LZgamma}
\end{center}
\end{figure}

We calculate the Landau-Zener probability at different dissipation rate $\gamma$, keeping other parameters fixed at $\Omega =0.3, V =5, v = 0.01$. For this, we initiate the system in the ground state at large negative $t$, and numerically solve the Shr\"odinger equation to extract the probability for the system to be in the excited state at large positive $t$. As $\gamma$ increases, the atoms dissipate faster to the ground state. It is evident from the Fig. \ref{LZgamma}. When dissipation is zero (Fig. \ref{LZgamma} (a)), a LZ transition occurs with infinite lifetime. With higher $\gamma$ values (from 0.001 to 0.01), the life time decreases gradually (Figs. \ref{LZgamma}(b)-(d)).

\subsection{Dependency on Rydberg Interaction }

Although Landau-Zener lifetime does not depend much on Rydberg interaction at high dissipation ($\gamma = 0.4$), there is a pronounced dependence at low  dissipation ($\gamma = 0.001$). We identify two distinct class of behavior in this connection. In Category-I, the Rabi-coupling $\Omega$ is moderate, and  in this range, a stronger $V$ enhances the LZ lifetime. In category-II, the  Rabi $\Omega$ is small, and increasing  $V$ here results a reduction of the LZ  lifetime. Between these two extreme limits, there lies an intermediate  $\Omega$ range: in which the LZ lifetime initially increases with $V$, reaches a maximum, and subsequently decreases with further increase in $V$.

\subsubsection{Category-I: moderate $\Omega$}
 
\begin{figure}
\begin{center}
\includegraphics[width=0.49\linewidth]{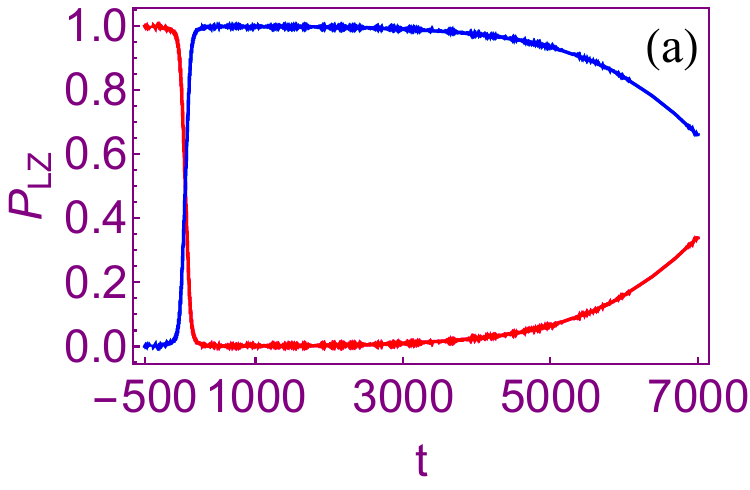}
\includegraphics[width=0.49\linewidth]{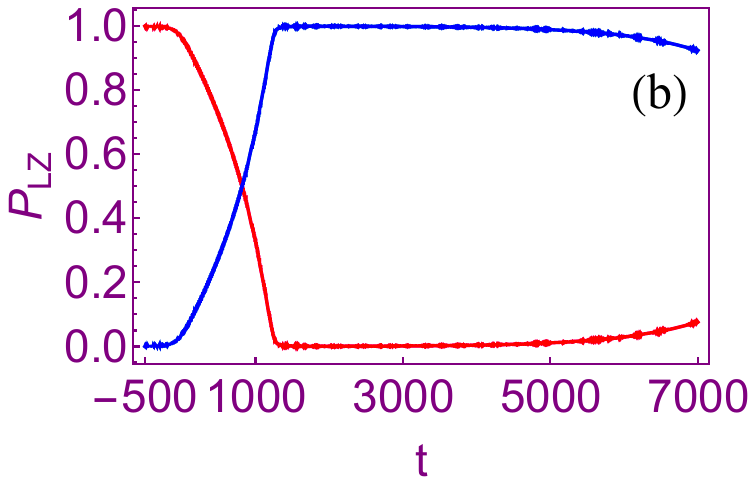}
\includegraphics[width=0.49\linewidth]{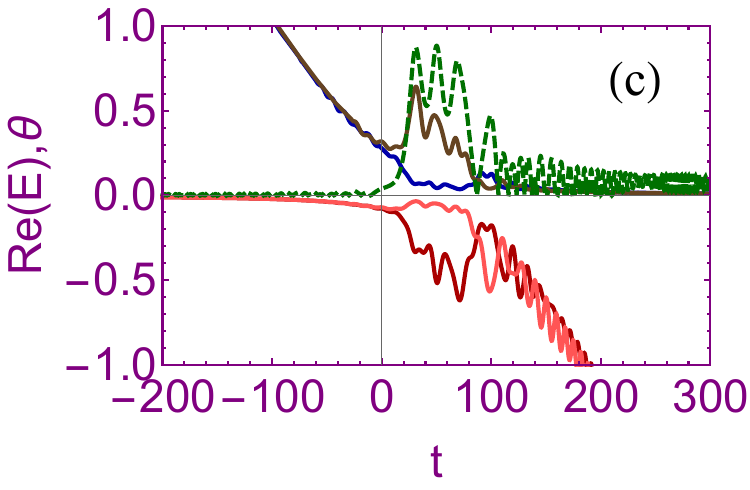}
\includegraphics[width=0.49\linewidth]{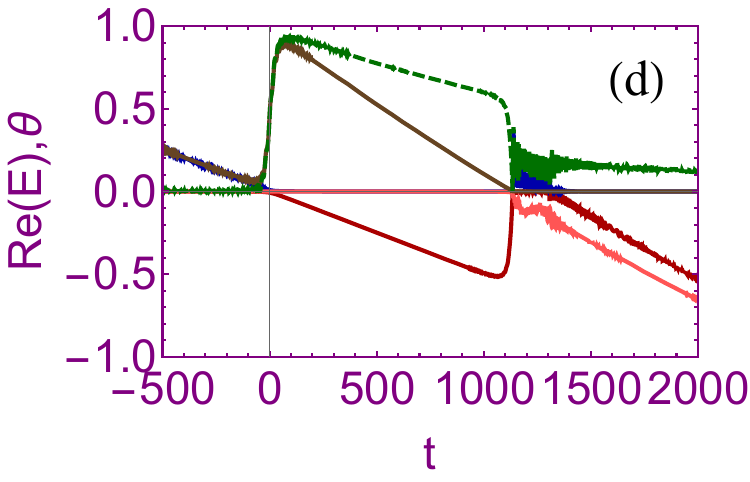}
\caption{ Category-I: Landau-Zener transition probability ($P{_\mathrm{LZ}}$) and eigen-energy spectra (Re(E)) of the non-Hermitian system at $\gamma = 0.001,\Omega =0.3, v = 0.01 $. $P{_\mathrm{LZ}}$ (red denotes occupation probability of the ground state, blue denotes the same for the excited state) for (a) $V=1$ and (b) $V=20$. Real part of energy eigen-spectra with time $t$, denoted by blue, brown, pink and maroon, solid lines, and for reference the order parameter $\theta$ (scaled by Rydberg strength $V$), denoted by dark green, dashed line is also plotted for (c)$V=1$ and (d)$V=20$.}
\label{cas1}
\end{center}
\end{figure}

 In this case, as Rydberg interaction $V$, increases, the $P_{\mathrm{LZ}}$ lifetime increases as well, as evident from  Figs. \ref{cas1}(a) and \ref{cas1}(b). Here, $\Omega$ (taken as $0.3$ in Fig. \ref{cas1}) sends sufficient number of atoms from the ground state to the Rydberg excited state, and directly competes with dissipation rate $\gamma$ that brings back the atoms to the ground state. As $V$ increases, the number of avoided crossing also increases. For example, in Fig. \ref{cas1}(c), there is a  single avoided crossing is there with $V=1$, whereas at $V=20$, there are (Fig. \ref{cas1}(d)) three distinct avoided crossings. Higher number of avoided crossings here enhances Landau-Zener transition, and prolongs the lifetime of LZ-excited states.

\subsubsection{Category-II : low $\Omega$}
\begin{figure}
\begin{center}
\includegraphics[width=0.49\linewidth]{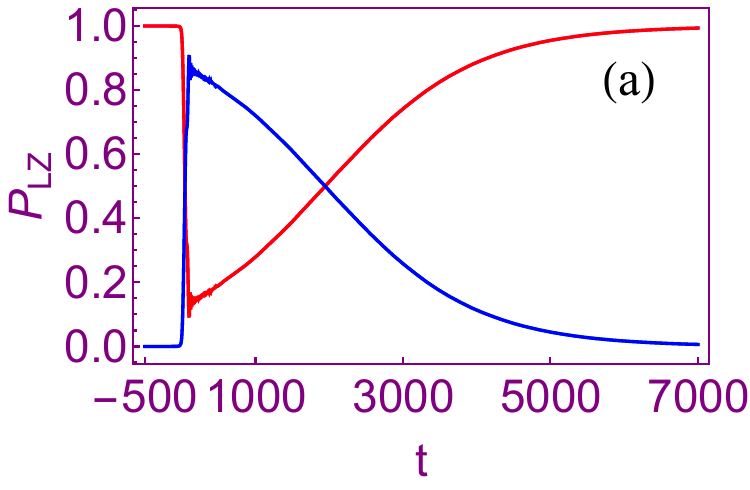}
\includegraphics[width=0.49\linewidth]{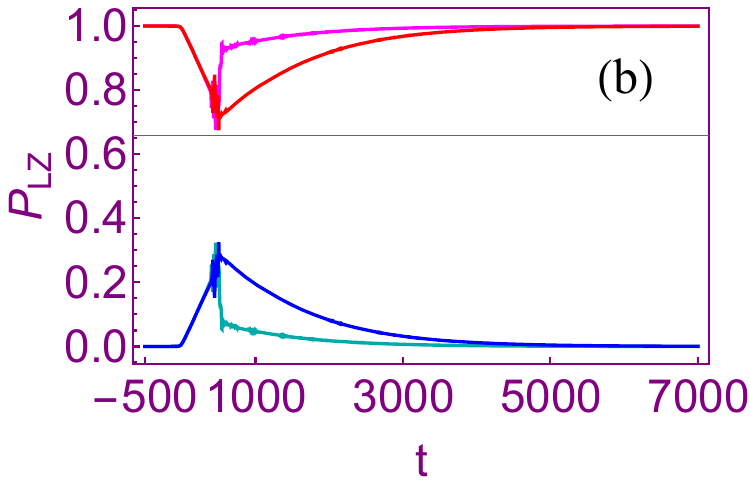}
\includegraphics[width=0.49\linewidth]{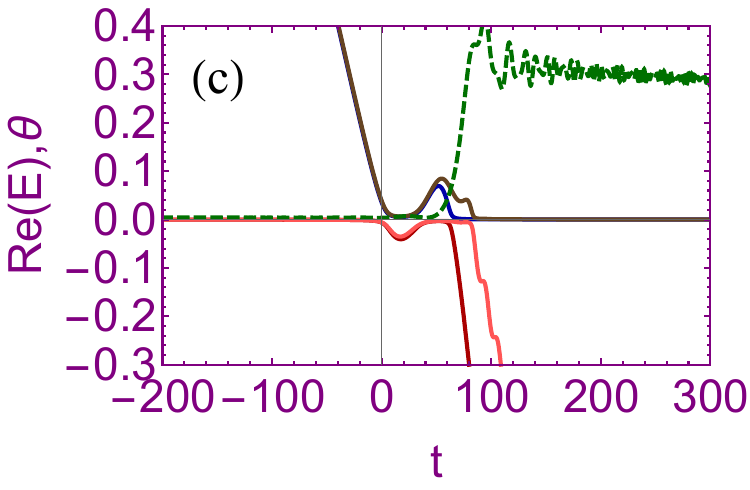}
\includegraphics[width=0.49\linewidth]{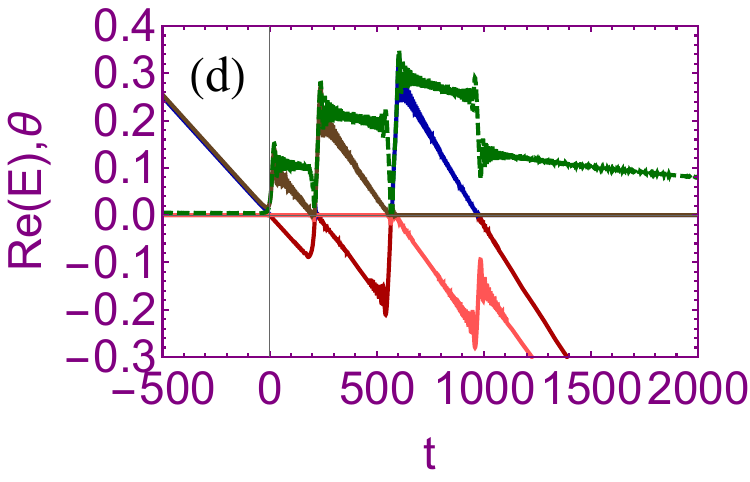}
\caption{ Category-II: Landau-Zener transition probability ($P{_\mathrm{LZ}}$) and eigen-energy spectra (Re(E)) of the non-Hermitian system at $\gamma = 0.001,\Omega =0.03, v = 0.01 $. $P{_\mathrm{LZ}}$ (red denotes occupation probability of the ground state , blue denotes the same for the excited state) for (a) $V=1$ and (b) $V=20$ . In the second plot,  $P{_\mathrm{LZ}}$ for the two sublattices does not overlap: red and magenta both denote the occupation probabilities of the ground state; blue and cyan both denote the same for the excited state). Real part of the energy eigen-spectra (blue, brown, pink and maroon, solid lines) are plotted with $t$, and the corresponding order parameter $\theta$ (scaled by Rydberg strength $V$), (dark green, dashed line) is shown for (c)$V=1$ and (d)$V=20$.}
\label{cas2}
\end{center}
\end{figure}

Here, Rabi coupling is taken to be lower (0.03 in Fig. \ref{cas2}), so, $\Omega$ is not enough to send sufficient number of atoms from the ground state to the excited state. Moreover, an increasing Rydberg interaction expands the gap between the ground and the excited state. These two incidents affect the Landau-Zener transition probability. As a result, as $V$ increases from 1 (Fig. \ref{cas2}(a)) to 20 (Fig. \ref{cas2}(b)), the $P_{\mathrm{LZ}}$ life time and $P_{\mathrm{LZ}}$ peak height diminishes gradually. On the other hand, at high $V$, the $P_{\mathrm{LZ}}$ life times of the two sublattices are not overlapped, these can be seen, distinctly in Fig. \ref{cas2}(b).


\section{ Dynamical Crossover In Landau-Zener Tunneling }
\label{crossover}

\begin{figure}
\begin{center}
\includegraphics[width=0.49\linewidth]{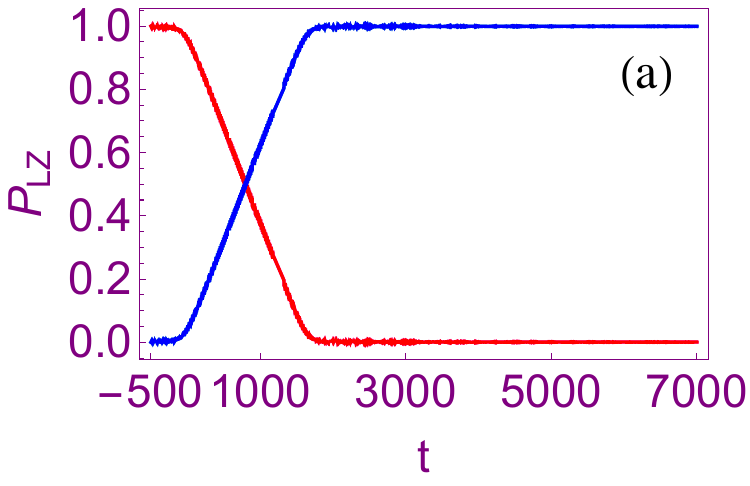}
\includegraphics[width=0.49\linewidth]{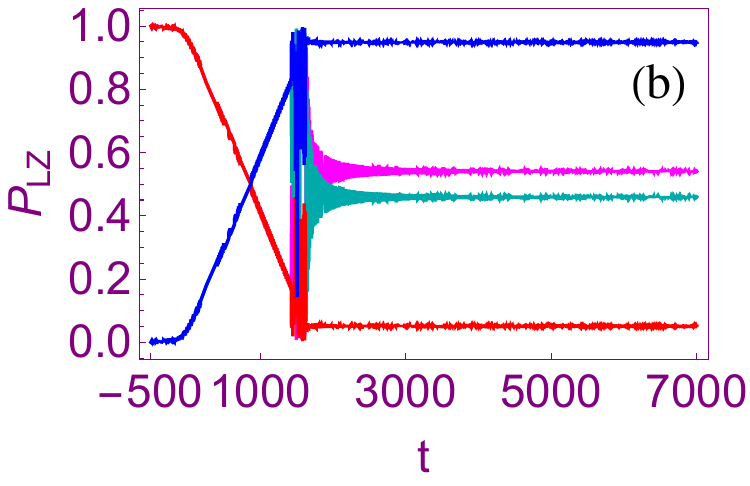}
\includegraphics[width=0.49\linewidth]{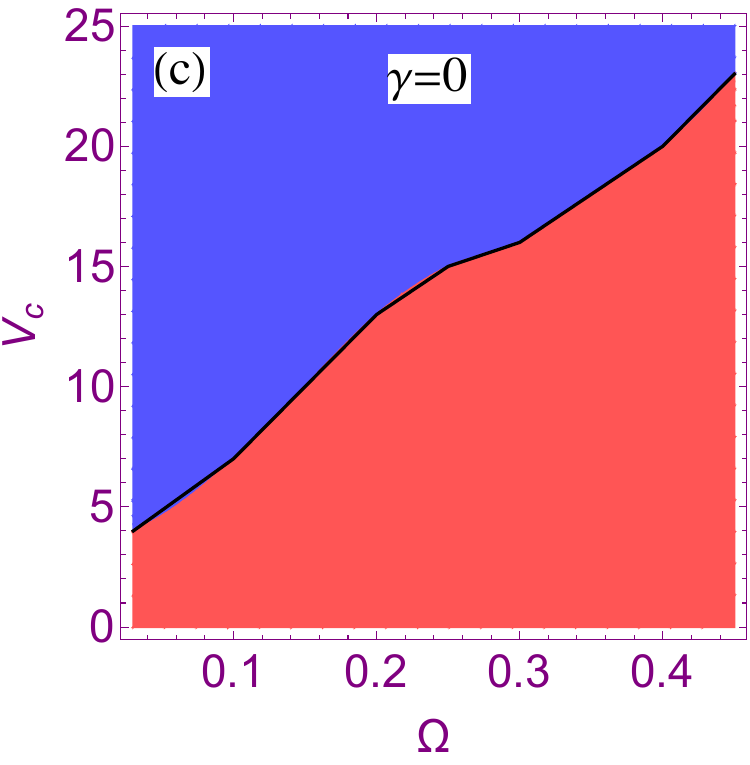}
\includegraphics[width=0.49\linewidth]{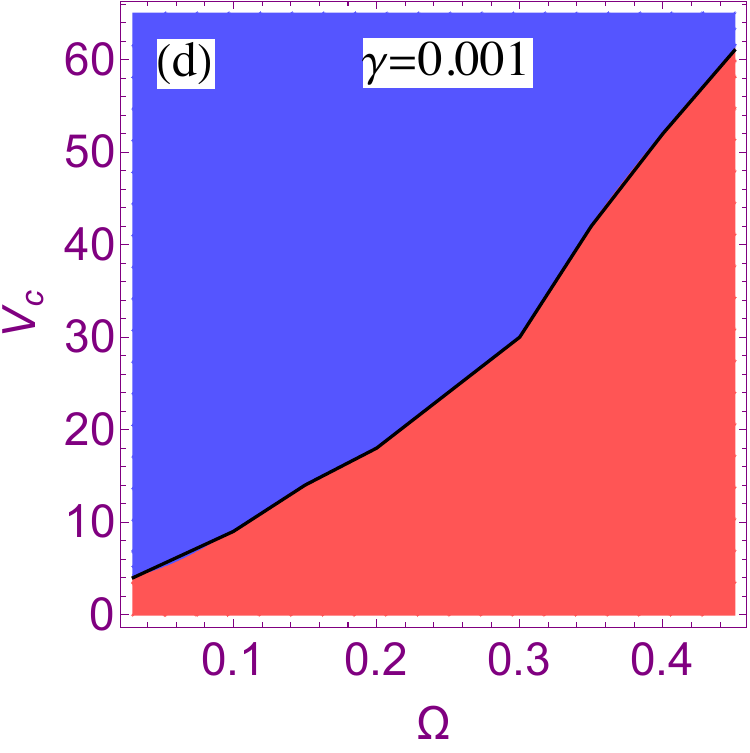}
\caption{ Dynamical crossover in Landau-Zener transition probability. Top panel: LZ transition probability at zero dissipation, with, $\Omega=0.3, v= 0.01$, (a) at critical Rydberg strength, $V_c=16$, where LZ probabilites (the ground state (red) and the excited state (blue)) of the two sublattices are same. (b) just above $V_c$, here, $V=17$, $P_{\mathrm{LZ}}$ is different for two sublattices (the ground states of the two sublattices are denoted by red and magenta colors and the excited states are denoted by blue and cyan colors respectively). Bottom panel: Critical Rydberg strength (black line) vs Rabi interaction plot, for (c) $\gamma=0$, (d) $\gamma=0.001$. Above the line (light blue), $P_{\mathrm{LZ}}$ of the two subllatices are different, below the line (light red), it is same. }
\label{Vcritical}
\end{center}
\end{figure}

Comparing Fig. \ref{cas1} and Fig. \ref{cas2}, we arrive at another interesting feature: in  category-I (moderate $\Omega$), both for small $V$ and large $V$, the Landau-Zener probability $P_{\scriptsize{\mbox{LZ}}}$ for the two sublattices match for all $t$, in spite of an AFM order being present in the order parameter plot. In contrast, in category-II (small $\Omega$),       
 $P_{\scriptsize{\mbox{LZ}}}$ of the two sublattices are same for small $V$ values, but they follow distinct paths over a considerable time-window at stronger blockade. 

This, actually, has a far deeper significance. This is an intrinsic property of the Rabi-coupled two-level Rydberg system, irrespective of whether dissipation is present or not. For example, in Fig. \ref{Vcritical}(a), $P_{\scriptsize{\mbox{LZ1}}}=P_{\scriptsize{\mbox{LZ2}}}$ for $\Omega=0.3$, $v=0.01$, $V=16$ in absence of dissipation. Then, if $V$ is increased to $17$ while keeping all other parameters same,  $P_{\scriptsize{\mbox{LZ1}}}$ and $P_{\scriptsize{\mbox{LZ2}}}$ separate out, as shown in Fig. \ref{Vcritical}(b). Thus, for $\Omega=0.3$, $v=0.01$, $V=16$ is a critical Rydberg strength that separates two types of Landau-Zener domains, and there is a dynamical crossover between the two.

This can be explained from the analytical expression for Landau-Zener excitation probability. For a simple two-level system with time dependent detuning, the Landau-Zener probability is given by \cite{niranjan2020landau, varghese2023maximally, melanathuru2022landau, glasbrenner2023landau, sun2025derivation}:

\begin{equation}
P_{\scriptsize{\mbox{LZ}}} =  \exp\Big({-{\pi\Omega^2}\over{2|v|}}\Big)  
\end{equation}

However, in our system, the effective detuning is not simply $\Delta = vt$, but
($-({{i \gamma}\over{2}}+ \Delta(t) )+V  \rho_{EE,2}$) (Eq. \ref{matrix}) for sublattice 1, and ($-({{i \gamma}\over{2}}+ \Delta(t) )+V  \rho_{EE,1}$) (Eq. \ref{matrix}) for sublattice 2. Therefore, $P_{\mathrm{LZ1}}$ can be equal to $P_{\mathrm{LZ2}}$ only if $\omega_1=\omega_2$ (Eq. \ref{rhoe}). 

As the system sweeps through the spectrum, it encounters multiple avoided crossings. At an avoided crossing, two adiabatic branches corresponding to a sublattice get strongly hybridized, leading to rapid redistribution of the state populations. This redistribution enables a transition between the uniform and the non-uniform configurations, and vice versa. Consequently, the order parameter $\theta$ switches from nonzero value to zero (or the opposite) near the avoided crossings in category-I (Figs. \ref{cas1}(c) and (d)), and for lower $V$ values in category-II  (Fig. \ref{cas2}(c)) implying $\omega_1=\omega_2$. As a result, $P_{\mathrm{LZ1}}=P_{\mathrm{LZ2}}$ throughout this region. Thus, despite the system exhibiting an AFM configuration between successive avoided crossings, the Landau-Zener excitation probabilities remain identical on the two sublattices, restoring $\mathcal{Z}_2$ symmetry at the crossing points. In this regime, the avoided crossings serve as spectral signatures of dynamical transition between the uniform and the non-uniform phases. 

In Regime II, however, for one or more of the AFM-to-uniform transitions, the avoided crossing does not coincide exactly with $\theta=0$. This regime corresponds to a larger ratio of $V/\Omega$, where the Rydberg blockade is stronger. Although rapid population redistribution still occurs near the avoided crossing, a residual AFM order persists (Fig. \ref{cas2}(d)) at that instant because of the stronger blockade. The AFM order then gradually decays as the sweep continues. Consequently, $\omega_1\neq\omega_2$ at that avoided crossing, and therefore, $P_{\mathrm{LZ1}}\neq P_{\mathrm{LZ2}}$. The $\mathcal{Z}_2$ sublattice symmetry at the avoided crossings are broken. 

Thus, an interaction-driven dynamical crossover separates two distinct Landau-Zener regimes. Fig. \ref{cas1}(b) belongs to regime I with equal Landau-Zener probabilities across the sublattices, while Fig. \ref{cas2}(b) belongs to regime II with a higher $V/\Omega$ value and sublattice-dependent Landau-Zener probabilities. These two regimes are shown in the bottom panel of Fig. \ref{Vcritical} (Fig. \ref{Vcritical}(c) for $\gamma=0$ and Fig. \ref{Vcritical}(d) for $\gamma=0.001$), where, the lower region (light red), represents regime I, with equal $P_{\mathrm{LZ}}$  and the upper region (light blue), indicates regime II, with sublattice dependent $P_{\mathrm{LZ}}$. These two regimes are separated by critical Rydberg strength $V_c$, as a function of Rabi coupling $\Omega$ (black lines in Figs. \ref{Vcritical}(c) and (d)).

\section{Conclusion}
\label{conclusion}

In this work, we study the effect of Rydberg interaction $V$ on the energy spectra and Landau-Zener transition in a dissipative Rydberg chain with a time-dependent detuning. First, we map our Hermitian Hamiltonian supplemented by dissipation through the Lindblad master equation, to an effective non-Hermitian Hamiltonian by neglecting quantum jumps.  We find that both Lindblad dynamics and the non-Hermitian formulation capture two distinct regions: a uniform one, and a non-uniform one in terms of the excited-state population distribution across the sublattices. By comparing these phase plots, we identify the region of validity of of this Lindblad to non-Hermitian mapping in the parameter space. 

We then extract the energy spectra of the effective non-Hermitian Hamiltonian and show that increasing the Rydberg interaction generates additional loops and branches in the real part of the eigen-energy spectra, signaling the emergence of antiferromagnetic order in the population distribution. We further find that the ratio $\Omega/\gamma = 1/2$ defines a boundary separating two qualitatively different spectral behaviors. For $\Omega/\gamma < 1/2$, the real part of the spectrum exhibits a true crossing, while the imaginary part shows an avoided crossing. For $\Omega/\gamma > 1/2$, this behavior is reversed.

Dissipation controls the lifetime of the Landau--Zener excited states. In the weak-dissipation regime, we find that the nature of the influence of the Rydberg interaction on this lifetime is sensitive to the Rabi coupling strength. For moderate Rabi coupling, increasing the Rydberg interaction enhances the Landau--Zener lifetime. In contrast, in the low $\Omega$ regime, stronger Rydberg interaction reduces the Landau-Zener lifetime.

Finally, we identify an intriguing dynamical crossover in the sublattice-dependent Landau--Zener probabilities. For a fixed Rabi coupling, there exists a crossover Rydberg interaction strength $V_c$, such that the Landau--Zener probabilities of the two sublattices are identical for $V \leq V_c$, and become distinct for $V > V_c$. This behavior arises because rapid and efficient redistribution of the sublattice populations occurs only in the vicinity of the avoided crossings, where the system undergoes a transition between uniform and non-uniform population distributions. Consequently, the antiferromagnetic order is generally confined to the region between two successive avoided crossings, and the system becomes nearly uniform in the immediate vicinity of an avoided crossing. For $V \leq V_c$, the sublattice population imbalance vanishes at the avoided crossing, resulting in identical Landau-Zener probabilities for the two sublattices. In contrast, for $V > V_c$, the stronger Rydberg blockade sustains a residual antiferromagnetic order even at the avoided crossing, leading to distinct Landau-Zener probabilities on the two sublattices.

\section{Acknowledgements}
R D would like to acknowledge Science and Engineering Research Board (SERB), currently, Anusandhan
National Research Foundation (ANRF), Department of Science and Technology, Govt. of India for
providing support under the CRG scheme (CRG/2022/007312), and Rashtriya Uchchatar Shiksha Abhiyan (RUSA) 2.0 (Ministry of Education, Govt. of India).
S I would like to acknowledge DST-INSPIRE, Department of Science and Technology, Govt.
of India, for providing support under the AORC scheme of the INSPIRE Program (DST/INSPIRE
Fellowship/2020/IF200534).
\bibliography{bibi1.bib}

\end{document}